# Bottom-up mechanism and improved contract net protocol for the dynamic task planning of heterogeneous Earth observation resources

Baoju Liu, Min Deng, Guohua Wu*, *Member, IEEE*, Xinyu Pei, Haifeng Li, Witold Pedrycz, *Fellow, IEEE*

*Abstract*—Earth observation resources are becoming increasingly indispensable in disaster relief, damage assessment and related domains. Many unpredicted factors, such as the change of observation task requirements, to the occurring of bad weather and resource failures, may cause the scheduled observation scheme to become infeasible. Therefore, it is crucial to be able to promptly and maybe frequently develop high-quality replanned observation schemes that minimize the effects on the scheduled tasks. A bottom-up distributed coordinated framework together with an improved contract net are proposed to facilitate the dynamic task replanning for heterogeneous Earth observation resources. This hierarchical framework consists of three levels, namely, neighboring resource coordination, single planning center coordination, and multiple planning center coordination. Observation tasks affected by unpredicted factors are assigned and treated along with a bottom-up route from resources to planning centers. This bottom-up distributed coordinated framework transfers part of the computing load to various nodes of the observation systems to allocate tasks more efficiently and robustly. To support the prompt assignment of large-scale tasks to proper Earth observation resources in dynamic environments, we propose a multiround combinatorial allocation (MCA) method. Moreover, a new float interval-based local search algorithm is proposed to obtain the promising planning scheme more quickly. The experiments demonstrate that the MCA method can achieve a better task completion rate for large-scale tasks with satisfactory time efficiency. It also demonstrates that this method can help to efficiently obtain replanning schemes based on original scheme in dynamic environments.

## I. INTRODUCTION

EARTH observation plays a vital role in many fields such as environment monitoring, disaster relief and urban analysis [1]. Earth observation resources are highly diversified, including satellites, unmanned aerial vehicles (UAVs), and airships. These resources have attracted an increasing attention of international organizations. For example, the U.S. Air Force Research Laboratory (AFRL) proposed the TechSat-21 program to verify the capability of coordinated work with multiple satellites [2, 3]. China also planned a program to achieve high quality Earth observation capabilities [4]. A high resolution Earth observation system is being built, including space-based, near space-based and air-based observation platforms. Observation resources are becoming smaller and more intelligent. Moreover, the development trend for Earth observation systems is to realize the coordinated usage of heterogeneous observation resources [5, 6].

The coordinated planning of heterogeneous observation resources is a hotspot in the field of Earth observation. Different observation tasks have different observation requirements in terms of resolution, time window, and sensing band. Earth observation users are in pursuit of more accurate services [7]. It is difficult for a single type of resource to meet the diversified requirements of various observation tasks. To this end, coordinated planning can take full advantage of various observation resources, and it can accomplish tasks that cannot be completed by a single type of resource.

It is necessary to perform task planning for observation resources dynamically. In general, observation environment involves considerable uncertainties in task execution process, such as signal anomalies, intense winds, thick clouds, and road congestion. In addition, the task requirements of observation targets, such as position and observation time, may change over time. Once an exception occurs, it needs to dynamically and promptly reformulate a reasonable observation scheme in accordance with current situation and original observation scheme.

It is difficult to reformulate a reasonable observation scheme efficiently in dynamically changing environments [8, 9]. With the current centralized planning strategy, the computing load of coordination mediator is too heavy, which may lead to the collapse of whole observation system if it fails. Moreover, the quality, timeliness of replanning scheme and the difference with original observation scheme are all factors that must be considered when reformulating observation scheme. The objective of coordinated task planning is to maximize the observation benefits of all resources while timely adjust the observation scheme if unpredictable disturbances occur [10].

The task planning has been extensively studied for each kind of observation resources. In the context of satellite scheduling, previous studies mainly focused on modeling techniques and solving algorithms. Many optimization models were extended and improved for satellite scheduling, such as task clustering [11], scene selection [12], knapsack problem optimization [13], and workshop scheduling [14]. Furthermore, different kinds of algorithms, including exact algorithm [11], heuristic algorithm [15] and metaheuristic algorithm [16, 17], have been used to solve this problem. For the task planning of air-based observation resources, many scholars studied the

This work was supported by the National Key Research and Development Program of China (No. 2016YFB0502600), Natural Science Fund for Distinguished Young Scholars of Hunan Province under Grant (No. 2019JJ20026), and the Fundamental Research Funds for the Central Universities of Central South University (No. 2018zzts198). (*Guohua Wu is the corresponding author.*)

Baoju Liu, Min Deng, Xinyu Pei and Haifeng Li are with the School of Geosciences and Info-Physics, Central South University, Changsha, 410000, China (e-mail: baojuliu@csu.edu.cn; dengmin@csu.edu.cn; 175011022@csu.edu.com; lihaifeng@csu.edu.cn).

Guohua Wu is with the School of Traffic and Transportation Engineering, Central South University, Changsha, 410000, China (e-mail: guohuawu@csu.edu.cn).

Witold Pedrycz is with the Department of Electrical & Computer Engineering, University of Alberta, Edmonton, AB T6R 2V4 Canada (e-mail: wpedrycz@ualberta.ca).

coordinated scheduling of multi-UAV. The task planning problem of multi-UAVs is essentially a task assignment problem with complicated constraints [18], which has usually been solved by optimization methods including mathematical programming [19], intelligent optimization [20, 18] and hierarchical decomposition [21].

It can be found that most task planning methods are oriented to a single type of resource and static environments, and under a centralized framework. For the coordinated task planning of multiple types of observation resources, several scholars also have proposed coordinated planning frameworks for heterogeneous resources [22, 23]. In addition, heuristic criteria [24, 25] and game theory [26] have been proposed to solve the coordinated planning problem of heterogeneous resources in a static environment. Although several researchers have studied the dynamic task planning problems of UAVs [27, 28], satellites [9] and other resources [8], the dynamic task planning of heterogeneous resources is still an unresolved problem to the best of our knowledge.

Existing studies proposed some coordinated planning frameworks and optimization algorithms in static environment. These coordinated planning frameworks first perform the task assignment process among planning centers. Then, the planning centers develop observation schemes for their resources. This top-down task planning process fails to take advantage of the distributed computing ability of different nodes of the observation systems. Moreover, the requirements for rapid replanning due to unexpected disturbances are ignored, which makes previous task planning methods inapplicable to actual observation scenarios. It is necessary to make the task planning of Earth observation resources adapt to the case of large-scale tasks in dynamic environments. In addition, it is needed to replace the centralized observation task allocation mode with a distributed task allocation mode, to increase the robustness and adaptation capability of the entire observation systems. This study analyzes and explores the task planning problems of heterogeneous Earth observation resources in dynamic and uncertain environments.

In this paper, a bottom-up distributed coordinated framework is proposed to realize the dynamic task planning of Earth observation resources. In contrast to the traditional top-down task assignment process, under this framework, once unpredictable disturbances occur, affected tasks are first tried to be replanned through the coordination among neighboring observation resources of the resource that is originally supposed to execute the task. Next, if the coordinated task replanning on the neighboring resource level fails, tasks are transferred to planning center, which then are tackled on the planning center level. In this manner, the tasks can be replanned more promptly in a dynamic environment.

In addition, a multiround combinatorial allocation (MCA) method is proposed to achieve rapid replanning of a large number of tasks. We integrate a contract net into the coordinated task planning process. With the MCA method, tender information is published by the resources that cannot perform tasks because of unpredictable disturbances. Multiple tasks can be combined and allocated to observation resources in one bidding process, which significantly improve the task allocation efficiency. Besides, a three-round task allocation process is designed to improve the task completion rate of allocation schemes. We further design a local search algorithm to select winning bidder in task allocation process. The experimental results demonstrate that the MCA method can deal with large-scale concurrent observation tasks dynamically and efficiently.

Our contribution made through this paper is threefold.

1) For the first time, we design a bottom-up distributed framework for the task replanning of heterogonous Earth observation resources. This approach is more consistent with dynamic and uncertain characteristics of the Earth observation environments.

2) We propose a dynamic environment oriented multiround combinatorial allocation (MCA) method to solve the large-scale task replanning problem rapidly. This method allocates tasks in a level by level manner based on the bottom-up distributed coordinated framework. In addition, task allocation process is combined with contract net. The experiments show that this method can significantly improve the efficiency and stability of task replanning.

3) We propose a local search algorithm to select winning bidder among observation resources and planning centers. This algorithm uses a probability parameter and a floating price interval mechanism to enhance the diversity of solutions and considerably improves the accuracy of solutions. At the same time, priority strategy is used to improve the convergence speed of the best solution.

The paper is organized as follows. In Section II, we review relevant works. In Section III, the coordinated task planning problem of Earth observation under dynamic environments is elaborated, and a bottom-up distributed coordinated framework is proposed. In Section IV, a multiround combinatorial allocation method with a solution algorithm is proposed to assign dynamic tasks. Section V presents and discusses experimental results. Section VI concludes this paper and presents directions for future work.

## II. RELATED WORK

In recent years, the construction and application of space-air-ground Earth observation systems have been carried out rapidly in several countries [29]. Many coordinated task planning methods for Earth observation resources have emerged according to the characteristics of different resources. With regard to the scheduling of a single type of observation resource, the research achievements of satellite scheduling are relatively fruitful. Many models, including scene selection [12], knapsack problem optimization [13], workshop scheduling [14], task clustering [9, 30], have been proposed to construct the mathematical models of satellite scheduling problem. Moreover, some intelligent optimization algorithms have been used to solve these models, such as particle swarm optimization [16], ant colony optimization [17] and genetic algorithm [31]. In addition, with the development trend of satellite constellation, coordinated task planning for multi-satellite has attracted an increasing attention. The TechSat-21 program [2, 3] COSMO-SkyMed constellation [32] and PLEIADES constellation [15] were launched to accomplish large scale tasks by using distributed multiple microsatellites.

Nag and Jiang et al. [33, 34] proposed coordinated task planning methods for multiple satellites from different perspectives. Besides, Xu, Tangpattanakul and Chang et al. [5, 6, 35] explored the task planning problem for agile satellites.

For the task planning of air-based observation resources, the coordinated scheduling of UAVs has received extensive attention. Many researchers treat UAVs as intelligent nodes and use consensus based [18], greedy-based [36] and MDP-based (Markov decision process) [37] methods to solve UAV scheduling problem. Moreover, intelligent algorithms have been widely used to optimize the task allocation scheme for UAVs, such as particle swarm optimization [16], genetic algorithms [38] and ant colony optimizing [17].

The capability of a single type of resource is often restricted by sensor band, endurance mileage, spatial resolution and other aspects when executing tasks. The coordinated task planning of multiple types of observation resources is an appealing method to conquer this bottleneck. Herold and Robinson explored the coordinated planning problem for space-air observation resources [25, 38, 40]. They proposed a coordinated framework for satellite and UAV resources and solved coordinated planning problem with linear programming method. Moreover, based on multiagent and game theory method, the task assignment and optimization problem of space-air resources was investigated by Li [23]. They further studied data transmission under an emergency environment [41]. Recently, Wu constructed a coordinated planning model for space-air-ground resources in accordance with four heuristic criteria and proposed an improved simulated annealing algorithm to solve the model [24]. These previous coordinated task planning methods can improve the complementary advantages of various observation platforms. In real applications (e.g., in the occurrence of disasters), however, real-time Earth observation services in uncertain environments are frequently and urgently required.

The literatures review above indicates that the independent planning of a single type of resource has been extensively investigated. These methods are mainly oriented to static environments, and they may fail in responding to emergency and dynamic situations for following reasons. First, the operation mode, running system, using rule and performance of heterogeneous resources in Earth observation network are significantly distinct and cannot be easily coordinated in a dynamic environment. Current top-down framework is not flexible and agile to be able to allocate tasks dynamically and efficiently. Second, real observation environments are highly dynamic and uncertain, while previous methods for heterogeneous resource planning cannot satisfy real-time task planning requirements. In particular, these methods are not suitable to dynamic task replanning which is required when unpredicted disturbances frequently occur. Third, top-down task planning mode fails to make full use of the increasingly intelligent resources which can potentially communicate and interact with each other.

In this paper, the task planning of Earth observation resources is extended from the static task planning of a single type of resource to the coordinated planning of multiple types of resources in dynamic environments. A bottom-up distributed framework is proposed in this paper, which can integrate heterogeneous resources and promptly allocate tasks dynamically through multilevel coordinated task planning framework. The essence of coordinated task planning problem is to allocate tasks to proper observation resources considering dynamic environment and complex constraints. Based on contract net protocol, a multiround combinatorial allocation method and an optimization algorithm are developed to reallocate the observation tasks affected by disturbances.

## III. PROBLEM DESCRIPTION AND ARCHITECTURE ANALYSIS

### A. Dynamic coordinated planning problem

Observation resources and tasks are two main types of objects in coordinated planning framework. Observation resources include multiple Space-Air-Ground observation platforms in general, such as satellites, airships, UAVs, monitoring vehicles etc. They are managed by different planning centers. Observation tasks are ground targets that need to be observed (points targets are specially considered in this study). It is generally difficult for an individual resource to complete large-scale observation tasks due to its limited capabilities. Besides, observation tasks may have different requirements involving spatial resolution, time window and spectrum. Therefore, coordinated task planning of multiple types of platforms is an inevitable trend in the development of Earth observation systems.

Coordinated task planning are periodically executed for Earth observation systems, according to the observation tasks submitted by users. Some scholars have proposed several heuristic methods to solve the coordinated task planning problem [24, 42]. They construct heuristic criteria to maximize observation benefits of observation schemes. Nevertheless, the requirements of observation tasks, observation resources and observation environments, may change during the coordinated task planning and task execution process. In this case, some scheduled tasks may not be finished anymore, and thus the existing planning scheme requires to be properly adjusted. This adjustment can be transformed into a dynamic multitask allocation problem in uncertain environments. In dynamic task allocation process, we must take into account several factors such as original observation scheme, resource load degree and task completion rate. Assume that the set of planning centers is denoted using $P = \{p_k | k=1, 2, ..., m\}$, where $m$ is the number of planning centers. Let $R_k = \{r_i | i= 1, 2, …, m_k\}$ be the set of observation resources managed by the observation center $p_k$, where $m_k$ is the number of resources. $T= \{t_j | j=1, 2, …, n\}$ is the set of tasks that need to be assigned, where $n$ is the number of tasks. The goal of this study is to dynamically allocate all tasks in $T$ to candidate resources to maximize observation benefits.

The difficulties in the dynamic task replanning among Earth observation resources lie in two aspects. First, the timely adjustment of task planning scheme is needed once task requirements, resources or environment change. It is difficult to regenerate a reasonable observation scheme efficiently which has high quality, high timeliness and a small difference from existing observation scheme. Second, the simultaneous adjustment of large-scale observation tasks will bring greater challenges to coordinated planning process. In traditional centralized planning mode, frequent coordinated task replanning almost occurs on one computing node (coordination mediator), the overload of this computing node will lead to the slow response of Earth observation system. Moreover, the robustness and adaptability of Earth observation system will be dramatically reduced once this node fails.

*B. Bottom-up distributed coordinated framework*

In general, each type of observation resource is managed by its own planning center in Earth observation system, as shown in Fig. 1, each planning center manages a single type of observation resource. The planning centers can communicate with each other, and some of the neighboring resources belonging to same planning center can communicate mutually.

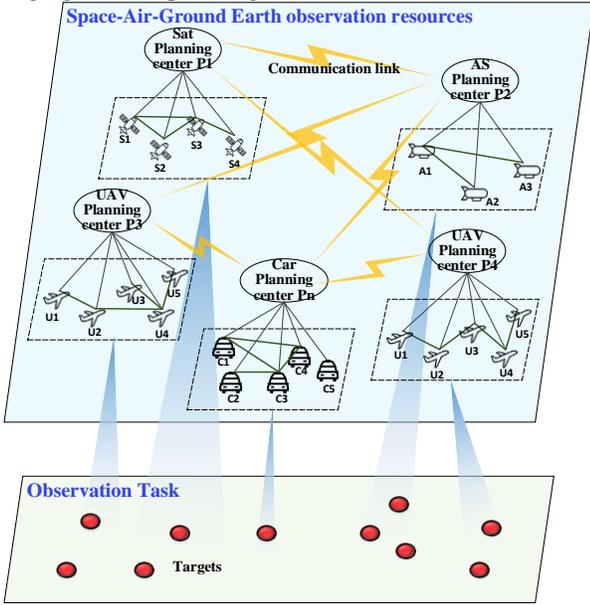

Fig. 1 Earth observation system.

Existing top-down task planning frameworks generally involve hierarchical structures. In these frameworks, coordination mediator first distributes tasks to each task planning center, which then develops observation schemes for its resources [22, 23]. This top-down planning framework works in a centralized mode, in which computing burden is mainly placed on single coordination mediator. Although this planning framework has advantages in getting globally better solutions, computational efficiency cannot be readily satisfied under emergency and dynamic conditions. Moreover, entire system would suffer a breakdown once coordination mediator cannot run normally. It may weaken the responsiveness and flexibility of overall observation system. Therefore, traditional top-down centralized task allocation mode is difficult to deal with dynamic task planning in uncertain environments.

This paper proposes a bottom-up distributed coordinated framework, which is compatible with existing resource communication mode. As shown in Fig. 2, this hierarchical framework consists of three levels, namely, neighboring resource coordination level, single planning center coordination level and multiple planning center coordination level.

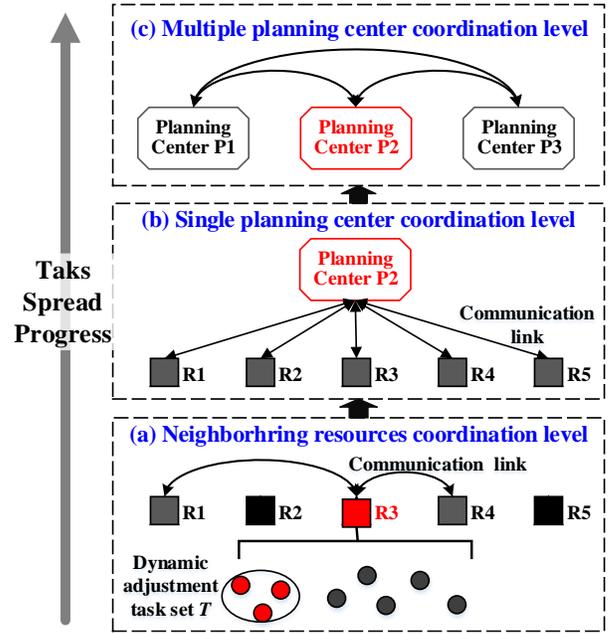

Fig. 2 Bottom-up distributed coordinated framework.

*1) Neighboring resource coordination level*

As shown in Fig. 2, in neighboring resource coordination level, if the tasks that have been scheduled to be executed by a resource $R3$ cannot be completed due to unexpected disturbances, $R3$ first try to coordinate with the communicable neighboring resources $R1$ and $R4$ to complete them. If these tasks cannot be cooperatively completed on this level, $R3$ passes task information and related observation requests to the higher level, i.e., the planning center $P2$.

*2) Single planning center coordination level*

After receiving task requirements, the planning center $P2$ negotiates with all observation resources that it manages, in order to develop an observation scheme. The remaining unfinished tasks are further passed to the upper level.

*3) Multiple planning center coordination level*

The planning center $P2$ negotiates with other planning centers to complete the remaining observation tasks. Other planning centers decide whether and how complete the tasks passed from planning center $P2$ according to their present own observation schemes and the capabilities of their internal resources, and then give feedback to $P2$. At this level, coordinated planning can be realized among multiple types of resources.

Every resource or planning center is a computing node in this bottom-up distributed coordinated framework, which greatly reduces the computational pressure of coordination mediator in top-down planning framework. Therefore, this framework would be more suitable to task planning in dynamic environment.

IV. MULTIROUND COMBINATORIAL ALLOCATION METHOD

Time efficiency is highly concerned in the dynamic task planning process, especially, it becomes a big challenge when a large number of tasks need to be dynamically replanned. Traditional contract net method can solve the problem of dynamic task allocation, and it is especially suitable to scenarios involving a single task, a single winning bidder, and a single round bidding [43]. However, it is difficult to meet the cases of large-scale tasks and high timeliness. Therefore, this paper proposes a multiround combinatorial allocation (MCA) method based on contract net to assign tasks through a three-round task allocation process. Especially, three round task allocation takes place at three different levels within the bottom-up distributed coordinated framework respectively. In each round task allocation, the improved negotiation process of the contract net is first designed to allocate tasks. Then, a local search algorithm is proposed to solve the combinatorial allocation scheme of multiple tasks, which is more efficient than single task auction algorithm. The MCA method supports task allocation with multiple tasks, multiple winning bidders, and multiround bidding.

*A. Contract net negotiation process*

As a high level communication interaction protocol, contract net supports the cooperation and competition of large scale resources in dynamic scenarios. The contract net transforms the passive allocation of tasks into the active bidding of resources by referring to "announce-bidding" mechanism employed in economic behavior. The negotiation process of contract net is improved to determine task allocation schemes in this paper. Two types of roles are involved in contract net: tenderer and bidder. Each resource or planning center can act as a tenderer or bidder. As shown in Fig. 3, in Earth observation system, the negotiation process of contract net can be divided into four stages: task announcement stage, bidding stage, awarding stage and task execution stage.

In the task announcement stage, once tasks submitted by users are detected, tenderer will publish tender announcement to candidate bidders. The message format of tender announcement is as follows: *TaskDocument=<ContractID, ContractType, TaskInfo, TaskRequirement, TaskWeight, ExpireTime, QuoteRequirement>*, where *ContractID* is the unique identifier of contracts. *ContractType* represents the type of contracts. *TaskInfo* is a detailed description of tasks, including task name, execution time, and spatial location. *TaskRequirement* indicates the additional constraint conditions of tasks, such as spectrum band and spatial resolution. *TaskWeight* represents the weight of tasks. *ExpireTime* is bid deadline. *QuoteRequirement* is the range of bidding prices.

In the bidding stage, candidate bidders submit their bidding document within expected time. After receiving tender announcements, bidders first evaluate the observation benefit obtained by performing tasks on the basis of contract requirements and existing observation schemes. Next, these bidders confirm bidding prices and return bidding documents to tenderer. The message format of bidding document is as follows: *BidDocument=<ContractID, Bid, ExecutionScheme, BidPrice, TaskSequences, IndicatorsStatus>*, where *Bid* is a Boolean variable indicating whether to bid for tasks. *ExecutionScheme* is the execution scheme of current task, including time, resolution, and spectrum. *BidPrice* represents the bidding price. *TaskSequences* is current task execution sequence. *IndicatorsStatus* is the completion status description of all task indicators and constraint condition.

In the awarding stage, after received all the bidding documents, a tenderer select the best task allocation scheme from all bids. The winning bidder is awarded contract. In this stage, the selection of the best allocation scheme can be considered as a winner determination problem (WDP). We designed a float interval-based local search (FLS) algorithm to solve this problem, which is described in subsequent sections.

In the executing stage, the tenderer sign contract with the winner. Then, this contract signing message is passed to all bidders. The winning bidder inserts new tasks into the task execution sequences.

This contract net negotiation process can take advantage of the distributed computing capability of observation resource and planning center. This distributed computing paradigm can significantly increase the efficiency of task allocation.

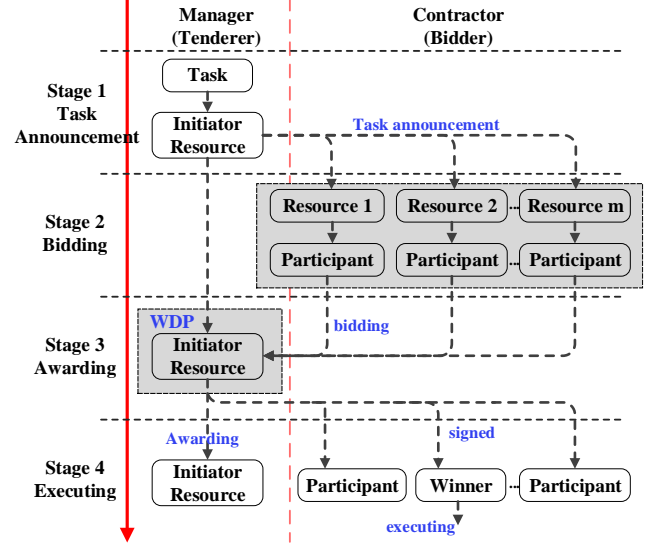

Fig. 3 Four stages of Earth observation system contract net.

*B. Multiround allocation based on the contract net*

In accordance with the bottom-up distributed coordinated framework, the MCA method is proposed to promptly allocate dynamic tasks. To overcome the task replanning problems caused by large scale tasks and dynamic environment, this method allocates tasks via three round negotiation process. At the first round, according to contract net, resource $r_i$ publishes tender information to its communicable neighboring resources. Next, the allocation scheme of combined tasks is determined through bidding documents feedback by neighboring resources. At the second round, a planning center receives unfinished tasks and negotiates with its resources to complete these tasks based on contract net. At the third round, the planning center coordinate with other planning centers jointly to formulate allocation scheme for remaining tasks. The algorithm can be described as follows.

Assume that $T=\{t_j|\ j=1, 2, \ldots, n\}$ is the set of tasks that need to be assigned, where $n$ is the number of tasks. Let $P = \{p_k|\ k=1, 2, \ldots, m\}$ be the set of planning centers, where $m$ is the number of planning centers, and planning centers can communicate among each other. The set of observation resources managed by planning center $p_k$ is denoted using $R_k = \{r_i|\ i=1, 2, \ldots, m_k\}$, where $m_k$ is the number of resources. $RS_i$ is the set of communicable neighboring resources that $r_i$ can communicate with. $GT = \{BT_e | e=1, 2, \ldots, bm\}$ is the collection of bidding tasks from all bidders, where $BT_e$ is the bidding task set from the $e$th bidder; $bm$ is the number of bidders; $GT \neq \phi$ and $GT \subseteq T$. $VT = \{V_e | e=1, 2, \ldots, bm\}$ is the bidding price set from bidders, $V_e$ is the bidding price of the $e$th bidder for bidding task set $BT_e$. $C_{best}^r[T]$ represents the tasks included in the winning bid $C_{best}^r$, $C_{best}^r[T] \subseteq T$.

Bidding price can be defined as the observation benefit obtained by performing tasks. It is related to the number of executable tasks, task weights, and consumption of resources. Based on this existing result [24, 42], we determine observation benefits as follows. Assume that $BT_e = \{bt_{ej} | j=1, 2, \ldots, n_e\}$ is the set of bidding tasks that bidder $b_e$ can complete, where $n_e$ is the number of tasks. Two types of heuristic criteria, including conflict degree and resource consumption degree, are proposed to measure the observation benefits of resource $b_e$ for executing bidding tasks.

The former denotes the degree to which other scheduled tasks (pending tasks) of resource $b_e$ cannot be completed along with task $bt_{ej}$ in $BT_e$ because of conflicts. The task is inclined to be allocated to the resource which has less conflicted scheduled tasks. Let $W_{ej} = \{w_k | k=1, 2, \ldots, m_{ejw}\}$ denote the weight set of scheduled tasks conflicting with $bt_{ej}$, where $m_{ejw}$ is the number of conflicting tasks. Then we can get the sum of weights of all conflicting tasks: $SW_{ej} = \sum_{k=1}^{m_{ejw}} w_k$. Therefore, the conflict degree $g_{ej}$ of resource $b_e$ executing task $bt_{ej}$ can be defined as

$$g_{ej} = \frac{SW_{ej}}{\max_{1 \leq j \leq n_e} SW_{ej}} \quad (1)$$

The latter is the degree that assigned tasks consume the remaining observation capability of this resource. The resource observation capability can be evaluated through its surplus observation duration, sensor-powering-on times, and flight distance. A task is inclined to be assigned to the resource that exhibits high observation capability. Suppose $D_{ej}$ is the required observation duration for resource $b_e$ executing task $bt_{ej}$. $D_e$ is the remaining observation duration of $b_e$. $R_{ej}$ is the required number of times for $b_e$ powering on its sensor to execute task $bt_{ej}$. $R_e$ is the remaining number of times for $b_e$ powering on its sensor. $cd_{ej}$ is the flight distance consumed by $b_e$ for observing target $bt_{ej}$. $L_e$ is the remaining fight distance of $b_e$. Moreover, the general model of resource consumption degree is described as

$$f_{ej} = \begin{cases} \left(\alpha \frac{D_{ej}}{D_e} + \beta \frac{R_{ej}}{R_e} + \gamma \frac{cd_{ej}}{L_e}\right)^2, & \text{if } D_{ej} \leq D_e, R_{ej} \leq R_e, cd_{ej} \leq L_e \\ 1, & \text{otherwise} \end{cases} \quad (2)$$

where coefficients $\alpha(0 \leq \alpha \leq 1)$, $\beta$ $(0 \leq \beta \leq 1)$, and $\gamma(0 \leq \gamma \leq 1)$, respectively, denote the weights of the three factors. The recommended value of $\alpha$, $\gamma$ and $\beta$ is 1/3 [24]. Therefore, overall bidding price $V_e$ of bidder $b_e$ for bidding task set $BT_e$ is

$$V_e = \sum_{j=1}^{n_e} \lambda_1 (1 - g_{ej}) + \lambda_2 (1 - f_{ej}) \quad (3)$$

where $\lambda_1$ and $\lambda_2$ are the weights of conflict degree and resource consumption degree, respectively. The recommended value of $\lambda_1$ and $\lambda_2$ is 0.5 [24].

---

**Algorithm 1** Multiround allocation algorithm

**Input**:
- $P$      Set of planning centers
- $T$      Task set
- $R_k$      Resource set managed by planning center $p_k$
- $RS_i$      Communicable neighboring resource set of $r_i$

**Output**:
- $C_{best}$      Task allocation scheme

1. **Begin**
2.    **If** $RS_i \neq \phi$ **then**
3.      Tenderer←$r_i$, Bidder←$RS_i$
4.      Public bidding
5.      All bidders submit bidding task collection $GT$ and bidding price set $VT$
6.      **Return** wining bidding scheme $C_{best}^r$ based on algorithm 2
7.      **If** $C_{best}^r[T] = T$ **then**
8.        $C_{best} \leftarrow C_{best}^r$
9.      **else**
10.        $T \leftarrow T - C_{best}^r[T]$
11.        Tenderer←$p_k$, Bidder←$R_k$
12.        Second round of bidding
13.        **Return** wining bidding scheme $C_{best}^{pIn}$
14.        **If** all tasks have been completed **then**
15.          $C_{best} \leftarrow C_{best}^r + C_{best}^{pIn}$
16.        **Else**
17.          $T \leftarrow T - C_{best}^r[T] - C_{best}^{pIn}[T]$
18.          Tenderer←$p_k$, Bidder←$P$-$p_k$
19.          Third round of bidding
20.          **Return** wining bidding scheme $C_{best}^p$
21.          $C_{best} \leftarrow C_{best}^r + C_{best}^{pIn} + C_{best}^p$
22.        **End if**
23.    **End if**
24. **Return** $C_{best}$
25. **End**

Algorithm 1 describes the process of multiround task allocation in detail. In the first round allocation, if resource $r_i$ has a nonempty communicable resource set: $RS_i \neq \phi$ (line 2), $r_i$ first publishes task announcement to $RS_i$; specifically, Tenderer=$r_i$, and Bidder=$RS_i$ (lines 3 and 4). As a bidder, each resource determines the task set $BT_e$ that can be completed and bidding price set $VT$. Next, these bidders submit $VT$ and the bidding task collection $GT$ which composed of $BT_e$ (line 5). Moreover, resource $r_i$ selects winning bidding scheme $C_{best}^r$ in accordance with the winning bidding determination algorithm presented as algorithm 2 (line 6). Due to the restrictions of resource capabilities, scheme $C_{best}^r$, however, usually cannot complete all tasks in $T$, that is $|C_{best}^r[T]| < |T|$ (line 7, line 9). Furthermore, in the second round allocation, current planning center $p_k$ acts as a tenderer, and resource set $R_k$ managed by observation center $p_k$ is bidder (line 11). The tenderers jointly negotiate winning bidding scheme $C_{best}^{pIn}$ for the remaining tasks in $T$ (lines 12 and 13). If there are still remaining unfinished tasks in the task set $T$ (line 17), the range of bidder is extended to all other planning center set $P-p_k$ (line 18). This process corresponds to the third round allocation (line 19). Moreover, planning center $p_k$ feeds back winning bidding scheme $C_{best}^p$ according to the algorithm 2 (line 20). Finally, based on the task allocation result of three rounds, final task planning scheme $C_{best}$ can be output (line 21).

*C. Winning bidder determination algorithm*

In the each round of negotiation process, all bidders submit bidding documents to *Tenderer*. Determination of winning bidders in each round determines the task allocation scheme. Different resources may observe same task in their feedback bidding scheme, thus there may be task conflicts among these bidding schemes. The best planning scheme can maximize observation benefits without any conflict. Therefore, an integer programming model is first constructed to describe optimization objectives and constraints. Furthermore, an algorithm is designed to select winning bidding scheme reasonably and promptly.

Definition: Conflict bidding task set. Assume that $BT_a$ and $BT_b$ are two bidding task sets from the $a$th and $b$th bidder, $BT_a \subseteq T$, $BT_b \subseteq T$ and $a \neq b$. If at least one task exists in two sets $BT_a$ and $BT_b$, namely, $BT_a \cap BT_b \neq \phi$, $BT_a$ and $BT_b$ are conflict bidding task sets. Otherwise, the two bidding sets are compatible. Thus, conflict matrix $M_{con}$, which is a symmetric matrix, can be constructed according to the conflict relationship between bidding task sets.

As shown in Eq (4), a subset of all bidding task collection $GT$ without conflicts can be selected as a feasible solution $C$, and the goal of Eq (4) is to maximize the sum of bidding prices in feasible solution $C$. The solution can be expressed by a Boolean set $X = \{x_e | e=1, 2, …, bm\}$, where $x_e=1$ denotes that bidding task set $BT_e$ is selected, $bm$ is the number of bidders. Therefore, objective function can be expressed as

$$\max \sum_{e=1}^{bm} V_e x_e \quad (4)$$

where $bm$ is the number of bidders, $V_e$ represents the bidding price of the $e$th bidder.

The constraints are

(a) $x_e \in \{0,1\}$;

(b) $x_i \otimes x_j = 0$  $1 \leq i \leq bm$, $1 \leq j \leq bm$, $i \neq j$.

where an operation $\otimes$ is defined as follows: if $x_i=1$, $x_j=1$, and $BT_i$ conflicts with $BT_j$, then $x_i \otimes x_j =1$; otherwise, $x_i \otimes x_j =0$. Constraint (b) indicates that each task can only be selected once, that is, the selected bidding task set does not conflict with others.

A float interval-based local search (FLS) algorithm (algorithm 2) is proposed in this paper to select winning bidding scheme in each round of bidding. The FLS algorithm searches for the best solution through multiple iterations. In general, the algorithm may become inefficient if it searches all candidate solution sets in each iteration. Therefore, to prevent the repeated searches of candidate solution set, a priority strategies were adopted, which can improve the convergence rate of the best solution. In the FLS algorithm, a priority search set $Q_B$ (line 2) was designed. $Q_B$ is the bidding task set compatible with the candidate best solution $C$, and there is no conflict between tasks in $Q_B$ and $C$. At the beginning of each iteration, the FLS algorithm searches $Q_B$ first and adds the bidding task with the highest price in $Q_B$ to candidate best solution $C$ (line 5). Moreover, as shown in Eq (5), temporary bidding task set *TemB* can be obtained according to candidate best solution $C$ (line 7).

| Algorithm 2 Float interval-based local search algorithm |
|---|
| **Input**: |
| $GT$     The collection of bidding tasks |
| $VT$     Bidding price set |
| $\rho$      Probability |
| $y$      Number of iterations |
| $\sigma$      Floating price interval |
| **Output**： |
| $C_{best}$    Best task allocation scheme |
| 1     **Begin** |
| 2     Initialize priority search set $Q_B \leftarrow GT$ |
| 3     **While** $C_{best}$ changes less than $y$ iteration **do** |
| 4        **If** $Q_B \neq \phi$ **then** |
| 5           $C \leftarrow C + max(Q_B)$ |
| 6        **Else** |
| 7           Candidate set $TemB \leftarrow GT - C$ |
| 8           **With probability $\rho$ do** |
| 9             Determines $V_{max}$ in *TemB* |
| 10         Select bidding task set $F_B$ from *TemB* within the floating price interval $\sigma$ of $V_{max}$. |
| 11          $B_{can} \leftarrow random(F_B)$ |
| 12         **Otherwise** |
| 13          $B_{can} \leftarrow random(TemB)$ |
| 14        **Done** |
| 15          $C \leftarrow C + B_{can}$ |
| 16        Update $C$: Remove bidding tasks that conflict with $B_{can}$ from $C$ |
| 17        **End if** |
| 18        **If** $\sum_{g=1}^{mc} Vc_g \geq \sum_{h=1}^{mb} Vb_h$ **then** |
| 19          $C_{best} \leftarrow C$ |
| 20        **End if** |
| 21        Update $Q_B$ according to conflict matrix $M_{con}$. |
| 22     **Done** |
| 23     **Return** $C_{best}$ |
| 24     **End** |

In addition, to avoid falling into local optimum in continuous optimization process, this algorithm designs a probability parameter $\rho$ (line 8) and a floating price interval $\sigma$ (line 10). Probability parameter is used to control the probability of selecting the bidding scheme with the highest price in iterative process. The floating price interval represents the gap between current bidding price and the maximum bidding price, which is used to determine the bidding price range of candidate bidding schemes.

The FLS algorithm first determines the maximum bidding price $V_{max}$ with probability $\rho$ (line 8 and 9), and then, as shown in Eq (6), this algorithm selects bidding task set $F_B$ from temporary bidding task set *TemB* within the floating price interval $\sigma$ of $V_{max}$ (line 10), that is, the difference between the bidding price of $F_B$ and the maximum bidding price $V_{max}$ is within interval $\sigma$. Furthermore, this algorithm randomly selects bidding task $B_{can}$ from set $F_B$ to add to candidate best solution $C$ (lines 11 and 15). The function *random (x)* is defined as randomly selecting an element from set $x$.

Alternatively, the FLS algorithm performs a random walk with a probability $1-\rho$. The algorithm selects bidding task $B_{can}$ randomly from $TemB$ (line 13) and adds it to candidate best solution $C$ with a probability $1-\rho$ (line 15). Consequently, the introduction of these random strategies partly facilitates enhancing the diversity of solutions.

$$TemB = B - C \quad (5)$$
$$F_B = \{B_i \mid |V_{max} - V_i| \le \sigma\}, B_i \in TemB \quad (6)$$

At the end of each iteration, candidate best solution $C$ is updated by judging the conflict relationship among bidding tasks (line 16). Moreover, to update the best solution via bidding price, let $VC = \{Vc_g \mid g=1, 2, \ldots, mc\}$ be the set of bidding price corresponding to the bidding task set in $C$. $VB = \{Vb_h \mid h=1, 2, \ldots, mb\}$ be the set of bidding price corresponding to the bidding task set in $C_{best}$. Therefore, the global best solution $C_{best}$ will be updated if $\sum_{g=1}^{mc} Vc_g \ge \sum_{h=1}^{mb} Vb_h$ (lines 18 and 19). Finally, the FLS algorithm updates $Q_B$ according to conflict matrix $M_{con}$ (line 21).

## V. Experimental Study

A simulation scenario was set up and the Aksu Prefecture (Xinjiang province, China) was selected as the experimental area in experiments. This experimental scenario was designed for emergency disasters in uncertain environments. We collected parameters of real Earth observation resources as basis for the resources performance in simulation scenario. To test the effectiveness of MCA method in different situations, two sets of experiments were designed as follows. The first set of experiments was used to verify whether the MCA method could effectively and promptly develop an observation scheme in the case of large scale tasks. The second set of experiments was used to verify whether the MCA method could make a reasonable observation scheme in dynamic environments.

### A. Comparative study of different task allocation methods for large scale tasks

As shown in Table 1, this scenario contained three different types of observation resources: satellites, UAVs and airships. These resources were managed by 4 planning centers (One satellite planning center managed 2 Earth observation satellites. 2 UAV planning centers managed 25 and 28 UAVs separately and one airship planning center managed 9 airships.). The performance parameters of all observation resources, including endurance, cruise speed, observation time and visible width, reflect the real capabilities of Earth observation resources. Moreover, in this experimental scenario, we formulated 5 groups of simulated tasks to reduce the impact caused by experimental random errors. Each group of data contained 600 randomly distributed point tasks that exceeded resource observation capability. And several attributes of tasks were set by referring to earthquake disaster statistics information. In addition, recommended input parameters for algorithms are also listed in Table 1.

Table 1 Parameter settings of the experiment A.

| Parameters | Value |
|---|---|
| Number of planning centers | 4 |
| Resource type in planning centers | {satellite, UAV, UAV, airship} |
| Number of resources in planning centers | {2, 25, 28, 9} |
| Visible width of satellite (m) | {8200,5000} |
| Side-swing angle of satellite (°) | {30,25} |
| Maximum observation duration of each satellite (s) | 2400 |
| UAV cruise speed (km/h) | {90,60} |
| Endurance mileage of UAV (km) | {21,30} |
| Visible width of UAV (m) | {500,600} |
| Maximum observation duration of each UAV (s) | 3000 |
| Cruise speed of airship (km/h) | 60 |
| Visible width of airship (m) | 650 |
| Maximum observation duration of each airship (s) | 4800 |
| Number of task groups | 5 |
| Number of tasks per group | 600 |
| Weight of task | Random values form 0 to 1 |
| Time window of task | Random values within 6 hours |
| Probability $\rho$ | 0.9 |
| Floating price interval $\sigma$ | 0.2 |
| Number of iterations $y$ | 10 |
| Number of clusters | 10 |

We compared the following five common coordinated planning methods with different number of tasks. The first method is sequential single item auction (SSA) method based on contract net [43]. In the SSA method, a single task is assigned to the bidder with highest bidding price by planning centers. Moreover, all the tasks are continuously assigned to resources item by item in accordance with contract net protocol. The second method is sorting allocation method according to the order of airship, UAV, satellite (AUS). This resource sequence can complete more tasks than any other sequence in single task allocation methods [24]. Both the SSA and AUS are single task sequence allocation methods, and they ignore the cooperation among the resources. The third and fourth methods are Mosek based coordinated planning (MCP) method and BnB based coordinated planning (BCP) method with a centralized framework, respectively. Their task assignment models can be expressed as Eq (4). The goal of coordinated planning is to maximize overall benefits. The Mosek optimization tools [44] are used as the solver to solve the integer programming model in the MCP method. The BCP method uses branch and bound algorithm to solve this integer programming model. The fifth method is task clustering allocation (TCA) method, which divides tasks into multiple clusters before assigning these tasks (*K-Means* method was selected as the clustering method in this experiment).

To verify the performance of MCA method for large scale tasks, we compared three important indicators for the five methods: task completion rate (TCR), consumed algorithm running time (CST) and average energy consumption (AEC). As shown in Eq (7), the average flight distance of observation resource to execute a single task was used to represent the AEC. A smaller average flight distance corresponded to a more reasonable planning scheme.

$$\bar{s} = \frac{\sum_{k=1}^{k=m}\sum_{i=1}^{i=m_k} S_{i,k}}{\sum_{k=1}^{k=m}\sum_{i=1}^{i=m_k} N_{i,k}} \quad (7)$$

where $S_{i,k}$ represents the sum of the flight distance of resource $r_i$ in planning center $P_k$. $N_{i,k}$ denotes the number of tasks that $r_i$ can complete.

The comparison results are shown in Fig. 4 and Fig. 5. Considering the cooperation among resources, MCP and BCP methods can complete more tasks than single task sequence allocation method. Their centralized planning framework, however, results in the low efficiency of both methods. Therefore, it is difficult for these methods to deal with large scale tasks. As shown in Fig. 4, the Mosek based coordinated planning (MCP) method has the highest task completion rate (TCR). In particular, the TCR remains 100% in the results that we can obtain (we cannot obtain results when the number of tasks is greater than 400 due to insufficient memory), and thus the approach appears to be a perfect planning method. The solution time of MCP method, however, becomes increasingly high with the increase in the number of tasks. The solution process may cost dozens of hours when 400 tasks are considered. Compared with its high task completion rate, the solution efficiency of MCP method renders it impractical, in particular, in the cases of emergency rescue. In contrast, the BCP method has a higher efficiency due to the characteristics of branch and bound algorithm (Fig. 5), although it is difficult to obtain a best solution. It can be found that the solving accuracy of BCP method involves a certain randomness due to the fluctuation (Fig. 4) of task completion rate.

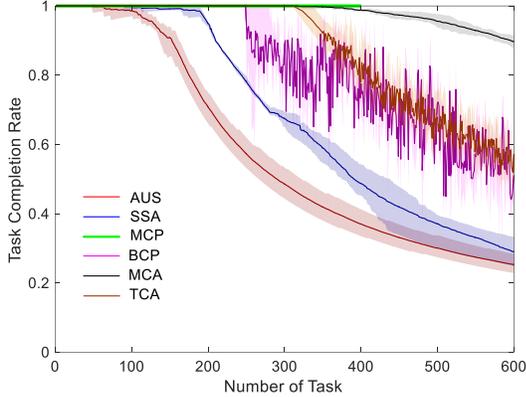

Fig. 4 Comparison result for task completion rate. The shadow area represents error.

The task completion rate and computational efficiency of SSA and AUS methods are quite similar because they both assign a single task one after another. The computational efficiency of these methods is significantly higher than that of MCP and BCP methods (Fig. 5). The algorithm running time of SSA and AUS methods is approximately proportional to the number of tasks, that is, $T_{method} \propto N_{task}$, where $T_{method}$ represents algorithm running time, and $N_{task}$ denotes the number of tasks. The SSA and AUS methods do not increase explosively when task scale is greater than 400 because they have basically lost task allocation ability (task completion rate decreased to about 30%). Without considering coordination mechanism, the task completion rate of SSA and AUS methods is considerably lower than that of MCP and BCP methods (Fig. 4). The SSA method selects the optimal solution according to the highest bidding price of all resources, while the AUS method considers only the profits of a single type of resource. Therefore, the task completion rate of SSA method is slightly higher than that of AUS method, although the algorithm of AUS method is more efficient.

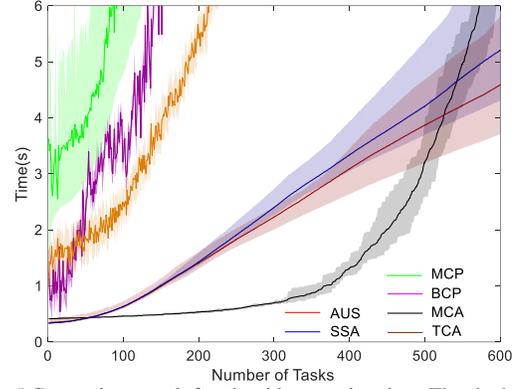

Fig. 5 Comparison result for algorithm running time. The shadow area represents error.

Overall, in general, a satisfactory completion rate and solution efficiency cannot be achieved simultaneously; the MCA method, however, realizes balances between them. The task completion rate of MCA method is quite similar to that of the MCP method, and it still achieves a completion rate of 95% when 500 tasks are considered (Fig. 4). The MCA method is significantly superior to the other three methods in terms of the number of tasks completed. The task allocation efficiency of MCA method is also profoundly superior to others before 400 tasks. After that, the running time grows rapidly because task scale has far exceeded the observation capability of resources. Those methods that are still effective in this stress test (task completion rate is greater than 0.5) are all increasing sharply in running time. The large-scale task stress test in the experiment, nevertheless, was designed to verify the performance of methods under unconventional extreme conditions, which rarely occur in reality.

In order to verify the efficiency of TCA method with different clusters, we analyzed the relationship between the number of clusters and the running time of TCA method. As shown in Fig. 6, the reduction in the number of task clusters leads to an increase in the number of tasks in each cluster, so more time-consuming it is to deal with task conflicts in these clusters. Essentially, to some extent, BCP method and SSA method can be regarded as two special cases in TCA method where the number of clusters $kc = 1$ and $kc = n$ ($n$ is the number of tasks). Experiment shows that the efficiency of TCA method is between BCP method and SSA method.

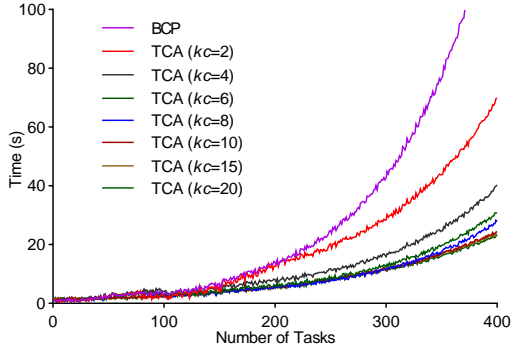

Fig. 6 Comparison result for different task clusters. $kc$ is the number of clusters.

To demonstrate the computational efficiency of MCA method in different levels, we decomposed the running time of MCA method into three levels according to bottom-up distributed coordinated framework. At the neighboring resource coordination level, tasks are assigned to communicable neighboring resources with a high efficiency. With the increase in the number of tasks, the current resources gradually expand the scope of tendering. The scale of the communication times in contract net is increasing gradually, which leads to a rapid decline in the time efficiency of MCA method (Fig. 7(a)). The task is passed to the second level once the observation capacity of neighboring resources reach their limitation (when the number of tasks increases to 250 in Fig. 7(b)). At the single planning center coordination level, the planning center calls for bids to internal resources. More dense dots in Fig. 7(b) indicates that more tasks need to be assigned on the second or third level. Therefore, running time increases, although it is still less than the time on the first level. At the multiple planning center coordination level, the planning center calls for bids to other planning centers. Since most of tasks have been completed on the first two levels, the third level takes less time to complete few remaining tasks. Overall, the first level takes the most time in the case of completing the most tasks. Nevertheless, the MCA method can maintain relatively high computational efficiency when the observation capability of resources meets task requirements (number of tasks <400).

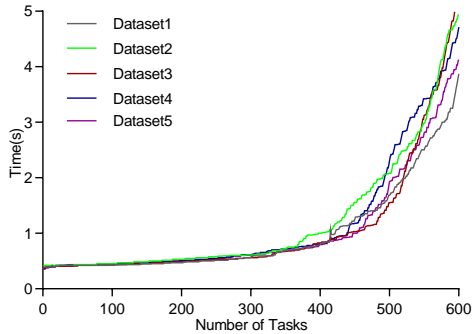

(a) Run time of algorithm for the resource coordination level.

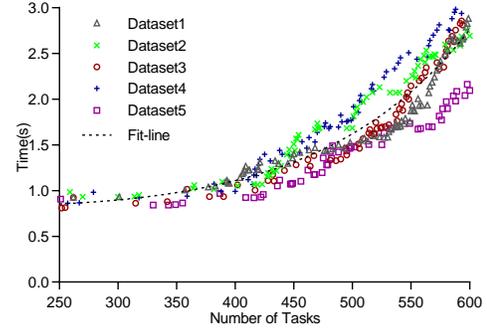

(b) Run time of algorithm for the single planning center coordination level.

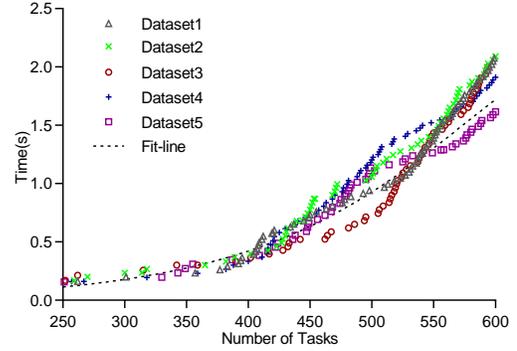

(c) Run time of algorithm for the multiple planning center coordination level.

Fig. 7 Run time of algorithm for the three levels within the MCA method.

Average energy consumption (AEC) denotes the quality of planning scheme. A smaller AEC means that less energy is consumed to complete a single task. The results of AEC have a high correlation with task completion rate. A higher task completion rate corresponds to a lower average energy consumption (Fig. 8). In addition, resource observation capacity can be used more rationally as the number of tasks increases. Therefore, the average energy consumption of executing a single task tends to reduce.

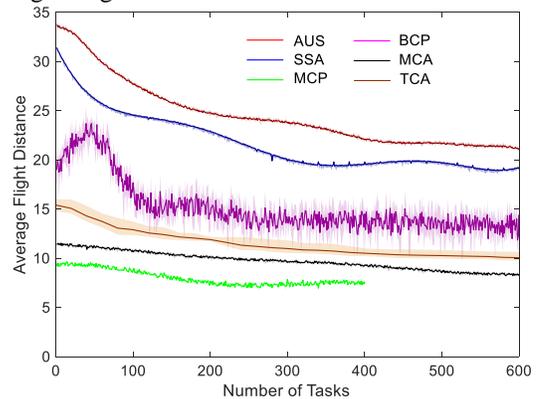

Fig. 8 Comparison result for average energy consumption. The shadow area represents error.

When we evaluate the task planning schemes derived from these methods, method efficiency, task completion rate and completion quality of planning schemes are equally important indicators. Consequently, considering these three indicators, the MCA method is the best choice for large-scale task allocation scenario.

## B. Comparative study of different task allocation methods in dynamic environments

We designed a set of experiments to verify the effect of the dynamic planning of MCA method. The outbreak of emergency disasters may lead to continuous dynamic changes and increases in tasks. Consequently, observation resources have to adjust planning schemes dynamically. To verify the dynamic planning effectiveness of MCA method in the case of insufficient observation capacity, experiment B reduced the amount of resources. As shown in Table 2, in this simulation scenario, 1 satellite, 18 UAVs and 3 airships were managed by 4 planning centers. These resources were initially required to observe 40 randomly distributed tasks. A total of 30 to 50 emergency tasks will appear dynamically while observation resources were executing tasks. The weight of each observation task is a random value between 0-1. The time window required to finish each task is distributed within 6 hours. Furthermore, the recommended parameters of algorithms are shown in the Table 2.

Table 2 Parameter settings of the experiment B.

| Parameters | Value |
|---|---|
| Number of planning centers | 4 |
| Resource type in planning centers | {satellite, UAV, UAV, airship} |
| Number of resources in planning centers | {1, 9, 9, 3} |
| Visible width of satellite (m) | 5000 |
| Side-swing angle of satellite (°) | 25 |
| Maximum observation duration of each satellite (s) | 2400 |
| UAV cruise speed (km/h) | {90,60} |
| Endurance mileage of UAV (km) | {21,30} |
| Visible width of UAV (m) | {500,600} |
| Maximum observation duration of each UAV (s) | 3000 |
| Cruise speed of airship (km/h) | 60 |
| Visible width of airship (m) | 650 |
| Maximum observation duration of each airship (s) | 4800 |
| Number of initial tasks | 40 |
| Number of dynamic planning | 5 |
| Number of increasing tasks each time | A random value from 30 to 50 |
| Weight of task | Random values from 0 to 1 |
| Time window of task | Random values within 6 hours |
| Probability $\rho$ | 0.9 |
| Floating price interval $\sigma$ | 0.2 |
| Number of iterations $y$ | 10 |

Observation resources reformulated new observation schemes according to different planning methods. The AUS, MCP and BCP methods developed the observation scheme for dynamic tasks by reclaiming unexecuted tasks and global replanning. Both SSA and MCA methods called for bid for dynamic tasks to all resources. Resources or planning centers feedback bidding price based on their current location and capabilities.

Task completion rate (TCR) is not the only relevant indicator in dynamic planning process. Replanning time (RPT) and rate of scheme change (RSC) are also key factors to determine the performance of planning method [45]. RSC is the ratio of the number of changed tasks, which are the tasks that are observed by different resources, to the number of all tasks in original scheme. A smaller RSC corresponds to less adjustment of resource actions. Similarly, a smaller RPT corresponds to a lower energy loss of resource and a better replanning scheme. Therefore, this paper compared these three indicators (TCR, RPT and RSC) of different methods in dynamic scenario.

In accordance with the experimental results presented in Table 3, the number of completed tasks of different methods continues increasing with the process of inserting new tasks dynamically. In contrast, task completion rate TCR tends to decline. Similar to the above experimental results, the TCR of MCP method is generally superior to that of other methods (Fig. 9). Single task sequence allocation methods (SSA and AUS) exhibit a disadvantage in terms of task completion rate. Task completion rate, however, is not the only indicator that we consider in dynamic planning scenario. In dynamic environments, planning schemes need to be developed as promptly as possible. Moreover, planning methods should minimize the impact of new planning scheme on original scheme. As shown in Fig. 10, the replanning time of different methods increases with the continuous insertion of tasks. The computational efficiency of MCP method is considerably lower than that of other methods. Running time is positively correlated with the total number of tasks (the dashed line in Fig. 10). This finding is consistent with the experimental conclusion obtained in previous sections. The MCA method exhibits notable advantages in terms of replanning time. Running time for 6 consecutive rounds of dynamic planning is always kept at a low level, which can meet the requirements for the dynamic observation of resources.

Table 3 Comparison of planning result in the process of dynamic replanning. NT denotes the number of new tasks inserted each time. AT represents the total number of all tasks.

| | NT | AT | MCP | | | MCA | | | BCP | | | SSA | | | AUS | | |
|---|---|---|---|---|---|---|---|---|---|---|---|---|---|---|---|---|---|
| | | | TCR | RPT(s) | RSC | TCR | RPT(s) | RSC | TCR | RPT(s) | RSC | TCR | RPT(s) | RSC | TCR | RPT(s) | RSC |
| 1 | 40 | 80 | 1 | 2.200 | 0.412 | 0.975 | 0.764 | 0.295 | 0.850 | 1.226 | 0.413 | 0.775 | 0.781 | 0.469 | 0.613 | 0.766 | 0.355 |
| 2 | 46 | 126 | 0.984 | 3.297 | 0.363 | 0.944 | 0.803 | 0.218 | 0.849 | 2.707 | 0.373 | 0.746 | 0.828 | 0.403 | 0.611 | 0.875 | 0.277 |
| 3 | 33 | 159 | 0.931 | 5.641 | 0.324 | 0.887 | 0.892 | 0.191 | 0.748 | 2.979 | 0.289 | 0.648 | 1.203 | 0.257 | 0.635 | 1.031 | 0.282 |
| 4 | 37 | 196 | 0.898 | 7.219 | 0.256 | 0.852 | 1.088 | 0.156 | 0.699 | 4.394 | 0.184 | 0.597 | 1.437 | 0.206 | 0.546 | 1.466 | 0.200 |
| 5 | 40 | 236 | 0.894 | 9.578 | 0.175 | 0.839 | 1.331 | 0.106 | 0.699 | 5.908 | 0.169 | 0.542 | 2.344 | 0.158 | 0.483 | 1.969 | 0.136 |
| 6 | 42 | 278 | 0.853 | 14.984 | 0.135 | 0.802 | 1.549 | 0.085 | 0.680 | 8.768 | 0.144 | 0.500 | 2.984 | 0.108 | 0.432 | 2.375 | 0.109 |

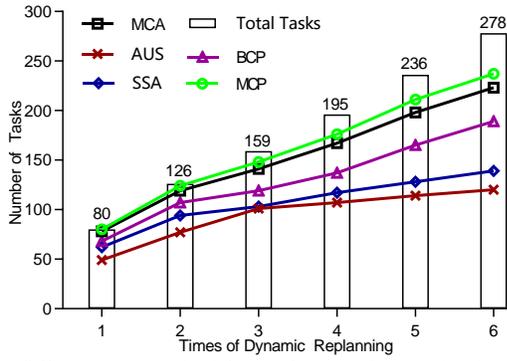

Fig. 9 Comparison result of the number of tasks completed in dynamic planning process. The histogram shows the total number of tasks, and the line graph shows the number of tasks completed.

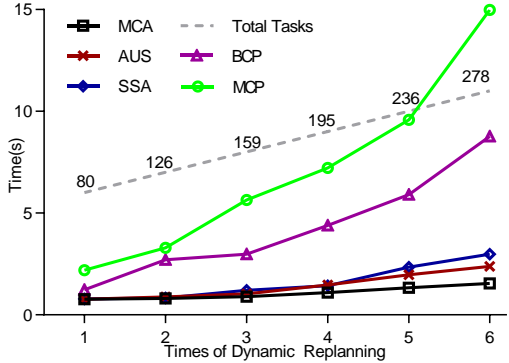

Fig. 10 Comparison result of replanning time. The dashed line represents the total number of tasks.

The rate of scheme change can reflect the impact of newly inserted tasks on original task. As shown in Fig. 11, the rate of scheme change of all methods gradually decreases as the number of new tasks increases. To a large extent, this is due to the fact that the total number of tasks (denominator) is gradually increased while the number of newly inserted tasks (numerator) is basically unchanged. Here, the ratio of the number of newly inserted tasks to the number of tasks in the last scheme before inserting tasks is used to indicate the occupancy rate (OR) of new tasks. The occupancy rate can indirectly reflect the impact of task cardinality on rate of scheme change. As shown in Fig. 11, the rate of scheme change is less than occupancy rate, which means that changed tasks in planning scheme are less than the newly inserted tasks. For example, in the round 3 of dynamic planning, the rates of scheme change of most methods are higher than values noted in histogram. This result indicates that in the planning scheme developed by methods MCP, BCP and AUS, the number of changed tasks is greater than the number of newly inserted tasks. Distinctly, for the MCA method, the rate of scheme change is always lower than task occupancy rate in the process of dynamic replanning, and is also lower than that of all other methods. Therefore, the planning scheme developed by the MCA method had a better scheme change rate than that of the other methods, whether or not the impact of task cardinality is eliminated.

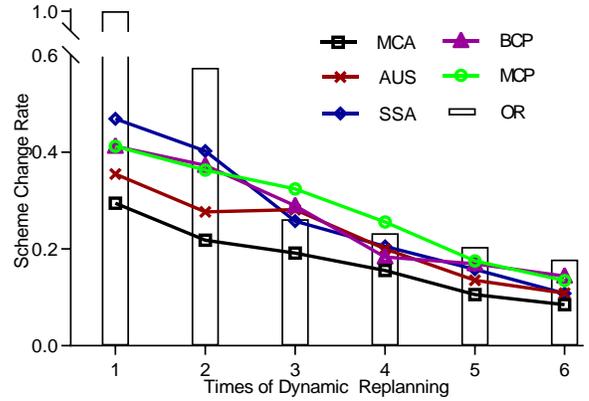

Fig. 11 Comparison result of scheme change rate. The histogram shows OR (the ratio of the number of new tasks to the number of existing tasks).

## VI. CONCLUSIONS

This paper proposes a bottom-up distributed coordinated framework, which embeds contract net into the coordinated planning process of Earth observation resources. According to this framework, we proposed a multiround combinatorial allocation (MCA) method by transforming the coordinated planning problem into a dynamic allocation problem. The MCA method considerably improves dynamic coordinated efficiency by performing multiple rounds of task allocation and contract net negotiation. Furthermore, the MCA method can promptly assign a large number of tasks through hierarchical allocation. In the process of task assignment, a key step is the determination of the best bid scheme. Therefore, this paper proposes a float interval-based local search (FLS) algorithm to solve winning bidding scheme selection problem in each round. The FLS algorithm uses priority strategies to improve the convergence speed of the best solution. The simulation experiments demonstrate the superiority of proposed method by performing large scale tasks and replanning test under a dynamic environment.

Earth observing resources, such as UAVs and satellites, have increasingly strong capabilities for information processing and intelligent computing. Therefore, in the future, we will focus on how observation resources actively discover and automatically track dynamic targets.

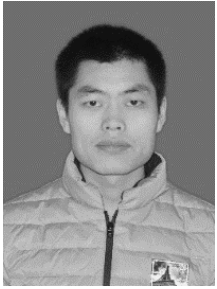

**Baoju Liu** received M.S. degree from the Central South University, Changsha, China, in 2017. He is currently pursuing his Ph.D. degree at the School of Geosciences and Info-Physics in Central South University, Changsha, China. His current research interests include Earth observation resources, dynamic programming and task assignment problem (TAP).

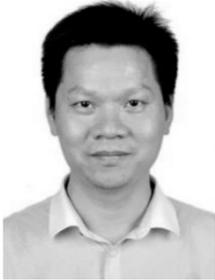

**Min Deng** received the Ph.D. degrees from Wuhan University in 2003 and the Asian Institute of Technology in 2004. He is currently the head of the Department of Geo-Information and the Doctoral Supervisor with the School of Geosciences and Info-Physics, Central South University. His current research interests include coordinated planning, spatio-temporal data mining, and spatio-temporal analysis and modeling. He ever hosted numerous major projects including a Key project of National Natural Science Foundation of China.

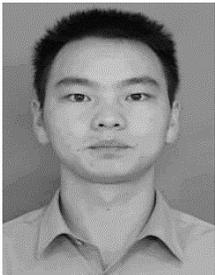

**Guohua Wu** received the B.S. degree in Information Systems and Ph.D. degrees in operations research from the National University of Defense Technology, Changsha, China, in 2008 and 2014, respectively. During 2012 to 2014, he was a Visiting Ph.D. Student with the University of Alberta, Edmonton, AB, Canada. From 2014 to 2017, he was a Lecturer at the College of Information Systems and Management, National University of Defense Technology, Changsha, China. He is now a Professor at the School of Traffic and Transportation Engineering, Central South University, Changsha 410075, China. His current research interests include planning and scheduling, evolutionary computation and machine learning. He has authored more than 50 refereed papers including those published in IEEE TSMC, IEEE TCYB and INS. He served as an Associate Editor of Swarm and Evolutionary Computation Journal and an editorial board member of International Journal of Bio-Inspired Computation.

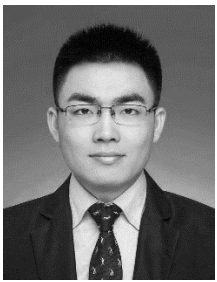

**Xinyu Pei** received the B.S. degree in management science and engineering in 2017 from the Central South University, Changsha, China. He is currently working toward the M.S. degree at the School of Geosciences and Info-Physics in Central South University, Changsha, China. His research interests include dynamic planning, and heuristic design.

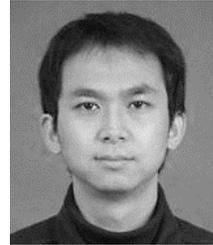

**Haifeng Li** (M'15) received the master's degree in transportation engineering from the South China University of Technology, Guangzhou, China, in 2005, and the Ph.D. degree in photogrammetry and remote sensing from Wuhan University, Wuhan, China, in 2009. He is currently a Professor with the School of Geosciences and Info-Physics, Central South University, Changsha, China. He was a Research Associate with the Department of Land Surveying and Geo-Informatics, Hong Kong Polytechnic University, Hong Kong, in 2011, and a Visiting Scholar with the University of Illinois at Urbana–Champaign, Urbana, IL, USA, from 2013 to 2014. His current research interests include geographic information services, spatial analysis, sparse representation, and machine learning. He has published over 30 journal papers.

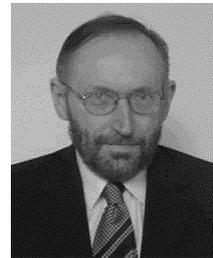

**Witold Pedrycz** (F'99) received the M.Sc., Ph.D., and D.Sc. degrees from the Silesian University of Technology, Gliwice, Poland. He is a Professor and Canada Research Chair (CRCComputational Intelligence) with the Department of Electrical and Computer Engineering, University of Alberta, Edmonton, AB, Canada. He is also with the Systems Research Institute, Polish Academy of Sciences, Warsaw, Poland. His main research interests include computational intelligence, fuzzy modeling and granular computing, knowledge discovery and data mining, fuzzy control, pattern recognition, knowledge-based neural networks, relational computing, and software engineering. He has published numerous papers in this area. He is intensively involved in editorial activities. He is the Editor-in-Chief of Information Sciences and serves as an Associate Editor of the IEEE Transactions on System Man Cybernetics: System and IEEE Transactions on Fuzzy Systems. He is also a Member of a number of editorial boards of other international journals.